\documentclass[aps,prb,twocolumn,floatfix,showpacs,10pt]{revtex4}
\pdfoutput=1 
\usepackage[english]{babel}
\usepackage{amsmath,amssymb}
\usepackage{graphicx}
\usepackage{color}
\usepackage{bm}
\usepackage{amssymb}
\usepackage{float}

\begin{document}
\title{Dislocation patterning in a 2D continuum theory of dislocations}

\author{Istv\'an Groma}
\email{groma@metal.elte.hu}
\affiliation{Department of Materials Physics, E\"otv\"os University Budapest,
H-1517 Budapest POB 32, Hungary}

\author{Michael Zaiser}
\affiliation{Institute for Materials Simulation, Department of Materials Science,
Friedrich-Alexander University Erlangen-N\"urnberg (FAU), Dr.-Mack-Str. 77, 90762
F\"urth, Germany}

\author{P\'eter Dus\'an Isp\'anovity}
\affiliation{Department of Materials Physics, E\"otv\"os University Budapest,
H-1517 Budapest POB 32, Hungary}

\begin{abstract}
Understanding the spontaneous emergence of dislocation patterns during plastic deformation is a long 
standing challenge in dislocation theory. During the past decades several phenomenological 
continuum models of dislocation patterning were proposed, but few of them (if any) are derived from 
microscopic considerations through systematic and controlled averaging procedures. 
In this paper we present a 2D continuum theory that is obtained by systematic averaging
of the equations of motion of discrete dislocations. It is shown 
that in the evolution equations of the dislocation densities diffusion like terms neglected 
in earlier considerations play a crucial role in the length scale selection of the dislocation 
density fluctuations.  It is also shown that the formulated continuum theory can be derived from
an averaged energy functional using the framework of phase field theories. However, in order 
to account for the flow stress one has in that case to introduce a nontrivial dislocation mobility 
function, which proves to be crucial for the instability leading to patterning. 
\end{abstract}

\maketitle

\section{Introduction}
Shortly after the first images of dislocations were seen in TEM it was realized that the 
dislocation distribution in a deformed crystalline material is practically never homogeneous. 
Depending on the slip geometry, the mode of loading and the temperature, rather different pattern 
morphologies (e.g. cell \cite{kavasaki1980}, labyrinth \cite{zhang2003}, vein \cite{siu2003} or 
wall \cite{ mughrabi1983} structures) emerge. There are, however, two important feature common to
all these patterns: It is almost always observed that the characteristic wavelength $\Lambda$ of 
the patterns  is proportional to the dislocation spacing, $\Lambda \propto 1/\sqrt{\rho}$ where 
$\rho$ is the total dislocation density, and inversely proportional to the stress at which the 
patterns have formed. These relationships are commonly referred to as ``law of 
similitude'' (for a general overview see Ref. \cite{kubin2011}). 

Since the early 1960s several theoretical and numerical attempts have been made to model  
dislocation pattern formation. The first models were based on analogies with other physical 
problems like spinodal decomposition \cite{holt1970}, patterning in chemical reaction-diffusion 
systems \cite{aifantis1985,aifantis2006}, internal energy minimization \cite{hansen1986}, or 
noise-induced phase transitions \cite{hahner1996,hahner1999}. Since, however, it is difficult 
to see how these models are related to the rather specific properties of dislocations 
(like long range scale free interactions, motion on well defined slip planes, or different types 
of short-range effects, etc.) they have to be considered as attempts to reproduce some
phenomenological aspects of the patterns based on heuristic analogies, rather than deriving them
from the physics of dislocation systems. 

To identify the key ingredients responsible for the emergence of inhomogeneous dislocation 
patterms, discrete dislocation dynamics (DDD) simulation is a promising possibility 
\cite{ghoniem1999,kubin1992,zbib1998,kubin2006}. The major difficulty, however, is that 
according to experimental observations the characteristic length scale of the dislocation patterns 
is an order of magnitude larger than the dislocation spacing \cite{kubin2011}. So, to be able to 
detect patterning in a DDD simulation one has to work with systems containing a large amount of 
dislocation lines: to see 3-5 pattern wavelengths in each spatial direction, the system size 
should be 30-50 times larger than $1/\sqrt{\rho}$. Especially in 3D this is a rather hard task.  
Although irregular clusters or veins are regularly observed in simulations 
\cite{devincre2001,devincre2002,hussein2015} clear evidence of the emergence of a 
characteristic length scale has not been published so far.           

During the past decade continuum theories of dislocations derived by rigorous homogenization of the 
evolution equations of individual dislocations have been proposed in 2D single slip 
\cite{groma1998,groma2001,groma2003,groma2007,groma2015} by the present authors and by Mesarovic {\it 
et.al.} \cite{mesarovic2010}. Later these models were extended to multiple slip by 
Limkumnerd and Van der Giessen \cite{limkumnerd2008}. In order to obtain closed sets of 
evolution equations for the dislocation densities, assumptions about the correlation properties of 
dislocation systems have to be made, but there are essential differences with earlier phenomenological 
models: Not only can these assumptions be shown to be consistent with the fundamental scaling properties of 
dislocation systems \cite{groma2001,zaiser2014}, but also the numerical parameters entering 
the theories can be deduced from DDD simulations in a systematic manner without fitting them 
in an ad-hoc manner to desired results. As a consequence, the models are predictive and can be directly 
validated by comparing their results to the outcomes of DDD simulations \cite{groma2003,groma2001,groma2004,groma2015}. 

Since 2D models are not able to account for several effects playing an important role in the 
evolution of the dislocation network, most importantly dislocation multiplication and junction formation, 
several 3D continuum theories have been proposed. For example Acharya \cite{acharia2007}  and later on 
Chen {\it et.al.} \cite{sethna2013} proposed models in which the coarse grained dislocation density (Nye's) 
tensor plays a central role. This tensor provides information about the distribution of 'geometrically necessary' 
dislocations with excess Burgers vector. However, incipient dislocation patterns are often associated with 
modulations in the total density of dislocations rather than modulations of the Burgers vector content. Hence it is 
doubtful whether models which concentrate on the transport of excess Burgers vector only can capture patterning.

Applying statistical approaches, El-Azab \cite{azab2000} and  Sedlacek {\it et.al.} \cite{sedlacek2007, kratochvil2008} 
suggested methods to handle curved dislocations, but the evolution equations obtained are valid only for quite specific 
situations. Considerable progress towards a generic statistical theory of dislocation motion in 3D 
has been made by Hochrainer {\it et.al.} \cite{hochrainer2007,hochrainer2014,hochrainer2015} by deriving a theory of 
dislocation density transport which applies to systems of three-dimensionally curved dislocations and can represent 
the evolution of generic dislocation systems comprising not only 'geometrically necessary' but also 'statistically stored'
dislocations with zero net Burgers vector.  Depending on the desired accuracy, the approach of Hochrainer allows to systematically 
derive density-based theories of increasing complexity. Recently this work was complemented by the 
derivation of matching energy functionals based upon averaging the elastic energy functionals of the corresponding discrete
dislocation systems \cite{zaiser2015}. In parallel, it was demonstrated how such energy functionals can be used to derive 
closed-form dislocation dynamics equations which are consistent not only with thermodynamics, but also with the constraints 
imposed by the ways in which dislocations move in 3D \cite{hochrainer2016}.

Concerning dislocation patterning, the general structure of continuum theories that is required for
predicting dislocation patterns that are compatible with the ``principle of similitude'' 
has been recently discussed by Zaiser and Sandfeld \cite{zaiser2014}. It was argued that no other length scales except
the dislocation spacing ($1/\sqrt{\rho}$) should appear in such theories - in other words, such theories
ought to be scale free. In Ref.\cite{zaiser2014} the authors also discussed a possible 
extension of the 2D continuum theory of Groma {\it et.al} \cite{groma2003} leading to the instability of the 
homogeneous dislocation density in a deforming crystal (details are discussed below). 

Some remarkable steps towards modeling pattern formation have been made by Kratochvil {\it et.al.} 
\cite{kratochvil2003} and recently in a 3D mean field theory by Xia and El-Azab \cite{elazab2015}. 
In order to obtain patterns, however, in both models specific microscopic dislocation 
mechanisms (sweeping narrow dipoles by moving curved dislocations, or cross slip, respectively) had 
to be invoked. In the present paper we adopt a more minimalistic approach where we consider no other 
mechanisms apart from the elastic interaction of dislocation lines. We analyze in detail the 
properties of a 2D single slip continuum theory of dislocations that is a generalization of the 
theory we have proposed earlier  
\cite{groma1998,groma2001,groma2003,groma2007,groma2015}. In the first part, the general structure 
of the dislocation field equations is outlined. To obtain a closed set of equations an assumption 
similar to the ``local density approximation'' often used for many-electron systems is used. After 
this, it is shown that the same evolution equations can be derived in a complementary manner, using
the formalism of phase field theories, from a functional which expresses the energy of the 
dislocation system as a functional of the dislocation densities. In the last 
part, by linear stability analysis of the trivial solution of the field equations the mechanisms 
for characteristic length scale selection in dislocation patterning are discussed.     

\section{Density based representation of a dislocation system: 
Linking micro- to mesoscale}
Let us consider a system of $N$ parallel edge dislocations with line vectors 
$\vec{l}=(0,0,1)$ 
and Burgers vectors $\vec{b}_{\pm}=\pm(b, 0,0)$. The force in the slip plane 
acting on a dislocation is $b \tau$ where $\tau$ is the sum of the shear stresses
generated by the other dislocations, and the stress $ \tau_{\rm ext} $ arising from 
external boundary displacements or tractions. It is commonly assumed that the velocity 
of a dislocation is proportional to the shear stress 
acting on  the dislocation (over-damped dynamics) \cite{groma1998,groma2003}. So, the 
equation of the motion of the $i$th dislocation positioned in the $xy$ plane at point $\vec{r}_i$ 
is  
\begin{equation}
\frac{dx_i}{dt}=M_0b\tau(\vec{r}_i)=M_0{b_i}\left(\sum_{j=1, j\ne i}^N s_j\tau_{\rm 
ind}(\vec{r}_i-\vec{r}_j)+\tau_{\rm ext} \right)
\label{eq_m}
\end{equation}
where $M_0$ is the dislocation mobility, $s_i=b_i/b=\pm 1$ is the sign of the $i$th 
dislocation (in the following often labeled '+' for $s= 1$ and '$-$' for $s=-1$), 
$\tau_{\rm ext}$ is the external stress, and  
\begin{equation}
\tau_{\rm ind}(\vec{r})=\frac{b\mu}{2\pi(1-\nu)}\frac{x(x^2-y^2)}{(x^2+y^2)^2}
\end{equation}
is the shear stress generated by a dislocation in an infinite medium. Here $\mu$ is the shear 
modulus and $\nu$ is Poisson's ratio.

After ensemble averaging as explained in detail elsewhere \cite{groma1998,groma2003,groma2007} one 
arrives at the following evolution equations:

\begin{eqnarray}
   \lefteqn{\partial_t \rho_+(\vec{r},t)
   +M_0b\partial_x 
\left[ \rho_+(\vec{r},t) \tau_{\rm ext} \right.} \label{eq:rho+} \\ &+ \left.\int \left\{
   \rho_{++}(\vec{r},\vec{r}',t)-\rho_{+-}(\vec{r},\vec{r}',t) \right \}
\,\tau_{\rm ind}(\vec{r}-\vec{r}'){\rm d}^2r'\right]=0,
 \nonumber 
\end{eqnarray}
\begin{eqnarray}
   \lefteqn{\partial_t \rho_-(\vec{r},t)
   -M_0b\partial_x 
\left[ \rho_-(\vec{r},t) \tau_{\rm ext} \right.} \label{eq:rho-} \\ &- \left.\int \left\{
   \rho_{--}(\vec{r},\vec{r}',t)-\rho_{-+}(\vec{r},\vec{r}',t) \right \}
\,\tau_{\rm ind}(\vec{r}-\vec{r}'){\rm d}^2r'\right]=0
 \nonumber 
\end{eqnarray}
where $\rho_{s}(\vec{r})$ and $\rho_{s,s'}(\vec{r},\vec{r}')$ with $s,s'\in\{+,-\}$
are the ensemble averaged one and two-particle dislocation density functions
corresponding to the signs indicated by the subscripts. It should be mentioned that Eqs. 
(\ref{eq:rho-},\ref{eq:rho+}) are exact, {\it i.e.} no assumptions have to be made to derive them, 
but certainly they do not represent a closed set of equations. In order to arrive at a closed 
set of equations one has to make some closure approximation to express the terms depending on the 
two particle density functions as functionals of the one particle densities (or one has to go to 
higher order densities.) The rest of this section is about suggesting a closure approximation 
consistent with discrete dislocation simulation results.      

For the further considerations it is useful to introduce the pair-correlation functions 
$d_{ss'}(\vec{r}_1,\vec{r}_2)$ defined by the relation
\begin{eqnarray}
\rho_{ss'}(\vec{r}_1,\vec{r}_{2},t)=\rho_{s}(\vec{r}_1)\rho_{s'}(\vec{r}_2)
(1+d_{ss'}(\vec{r}_1,\vec{r}_{2})) 
\label{eq:hd0}
\end{eqnarray}
According to DDD simulations the pair-correlation functions defined above decay to zero within a
few dislocation spacings \cite{groma2001}. As a result of this,
if the total dislocation density $\rho=\rho_++\rho_-$  varies slowly enough in
space, we can assume that the correlation functions  depend explicitly only on
the relative coordinate $\vec{r}_1-\vec{r}_{2}$, see Refs. \cite{groma2003,groma2007}. 
The direct $\vec{r}_1$ (or $\vec{r}_{2}$)  dependence appears only through the local 
dislocation density, i.e.,
\begin{eqnarray}
d_{ss'}(\vec{r}_1,\vec{r}_{2})=d_{ss'}(\vec{r}_1-\vec{r}_{2},\rho(\vec{r}_1)).
\label{eq:d0}
\end{eqnarray} 
(Since  $d_{ss'}$ is short ranged  in  $\vec{r}_1-\vec{r}_{2}$, it does not
make any difference if in the above expression $\rho(\vec{r}_1)$ is replaced by
$\rho(\vec{r}_2)$.) 
 
In the case of a weakly polarized dislocation arrangement where 
$\rho_+-\rho_-\ll\rho$, the only relevant length scale is the average 
dislocation spacing. So, for dimensionality reasons the $\rho$ dependence of 
$d_{ss'}$  has to be of the form
\begin{eqnarray}
d_{ss'}(\vec{r}_1,\vec{r}_{2})=d_{ss'}((\vec{r}_1-\vec{r}_{2})\sqrt{\rho(\vec{r}
_1)}).
\label{eq:drho}
\end{eqnarray}

By substituting Eq. (\ref{eq:hd0}) into Eqs. (\ref{eq:rho+},\ref{eq:rho-}) and introducing the 
GND dislocation density $\kappa=\rho_+-\rho_-$ one arrives at
\begin{eqnarray}
   \partial_t \rho_+(\vec{r},t) +M_0b\partial_x 
\left\{\rho_+\left[ \tau_{\rm ext}+\tau_{\rm sc}+\tau_+ \right]\right\}=0 \label{eq:rhod+}
\end{eqnarray}
\begin{eqnarray}
   \partial_t \rho_-(\vec{r},t) -M_0b\partial_x 
\left\{\rho_-\left[ \tau_{\rm ext}+\tau_{\rm sc}+\tau_- \right]\right\}=0 \label{eq:rhod-}
\end{eqnarray}
where
\begin{eqnarray}
\tau_{\rm sc}(\vec{r})=\int \tau_{\rm ind}(\vec{r}-\vec{r}')\kappa(\vec{r}'){\rm d}^2r',
\label{eq:tausc}
\end{eqnarray}
commonly called the ``self consistent'' or ``mean field'' stress, is a non-local functional
of the GND density, whereas the stresses  
\begin{eqnarray}
 \lefteqn{\tau_+(\vec{r})=\int \left[ \rho_+(\vec{r}')d_{++}(\vec{r}-\vec{r}') \right.} 
\nonumber \\&- \left. \rho_-(\vec{r}')d_{+-}(\vec{r}-\vec{r}')  \right] \tau_{\rm 
ind}(\vec{r}-\vec{r}') {\rm d}^2r', \label{eq:taud+}
\end{eqnarray}
and 
\begin{eqnarray}
 \lefteqn{\tau_-(\vec{r})=-\int \left[ \rho_-(\vec{r}')d_{--}(\vec{r}-\vec{r}') \right.} 
\nonumber \\&- \left. \rho_+(\vec{r}')d_{-+}(\vec{r}-\vec{r}')  \right] \tau_{\rm 
ind}(\vec{r}-\vec{r}') {\rm d}^2r' \label{eq:taud-}
\end{eqnarray}
depend on dislocation-dislocation correlations. Finally let us introduce the quantities
\begin{eqnarray}
 \tau_v&=&\frac{\tau_++\tau_-}{2} \label{eq:tauv},\\
 \tau_a&=&\frac{\tau_+-\tau_-}{2} \label{eq:taua}.
\end{eqnarray}
With these quantities Eqs. (\ref{eq:rhod+},\ref{eq:rhod-}) read 
\begin{eqnarray}
   \partial_t \rho_+(\vec{r},t) +M_0b\partial_x 
\left\{\rho_+\left[ \tau_{\rm ext}+\tau_{\rm sc}+\tau_v+\tau_a \right]\right\}=0 
\label{eq:rhodd+},\\
   \partial_t \rho_-(\vec{r},t) -M_0b\partial_x 
\left\{\rho_-\left[ \tau_{\rm ext}+\tau_{\rm sc}+\tau_v-\tau_a \right]\right\}=0 \label{eq:rhodd-}.
\end{eqnarray}
In explicit form the stresses  $\tau_v$ and $\tau_a$ are given by
\begin{eqnarray}
 \tau_v(\vec{r})=&\int \left[ \rho(\vec{r}')d_a(\vec{r}-\vec{r}') 
+  \kappa(\vec{r}')d_s(\vec{r}-\vec{r}')  \right] \nonumber \\&
* \tau_{\rm ind}(\vec{r}-\vec{r}') {\rm d}^2r' \label{eq:taudv},\\
 \tau_a(\vec{r})=&\int \left[ \rho(\vec{r}')d_p(\vec{r}-\vec{r}') 
+  \kappa(\vec{r}')d_{a'}(\vec{r}-\vec{r}')  \right] \nonumber \\&
* \tau_{\rm ind}(\vec{r}-\vec{r}') {\rm d}^2r' \label{eq:tauda},
\end{eqnarray}
with
\begin{eqnarray}
 d_s&=\frac{1}{2}(d_{++}+d_{--}+d_{+-}+d_{-+}), \\
 d_p&=\frac{1}{2}(d_{++}+d_{--}-d_{+-}-d_{-+}), \\
 d_a&=\frac{1}{2}(d_{++}-d_{--}-d_{+-}+d_{-+}), \\
 d_{a'}&=\frac{1}{2}(d_{++}-d_{--}+d_{+-}-d_{-+}).
\end{eqnarray}

We note some symmetry properties of the pair correlation functions: (i) the functions $d_{++}$ and 
$d_{--}$ must be invariant under a swap of the two dislocations and thus represent 
even functions of $\vec{r}$; (ii) for dislocations with different signs one gets from the 
definition of correlation functions that $d_{+-}(\vec{r})$=$d_{-+}(-\vec{r})$. Hence $d_s(\vec{r})$ 
and $d_p(\vec{r})$ are even functions, while the difference $d_{+-}-d_{-+}$ appearing in $d_a$ and 
$d_{a'}$ is an odd function. 

For the further considerations it is useful to introduce the notations
\begin{eqnarray}
 \tau_f(\vec{r})=-\int  \rho(\vec{r}')d_a(\vec{r}-\vec{r}') \tau_{\rm 
ind}(\vec{r}-\vec{r}') {\rm d}^2r' \label{eq:tauf}
\end{eqnarray}
referred to ``friction stress'' hereafter,
\begin{eqnarray}
 \tau_b (\vec{r})= \int\kappa(\vec{r}')d_s(\vec{r}-\vec{r}') \tau_{\rm 
ind}(\vec{r}-\vec{r}') {\rm d}^2r' \label{eq:taud}
\end{eqnarray}
commonly called ``back stress'', 
\begin{eqnarray}
\tau_d(\vec{r})=\int  \rho(\vec{r}')d_p(\vec{r}-\vec{r}') \tau_{\rm ind}(\vec{r}-\vec{r}') {\rm 
d}^2r' \label{eq:taua1}
\end{eqnarray}
called ``diffusion stress'', and
\begin{eqnarray}
 \tau'_f(\vec{r})=\int   \kappa(\vec{r}')d_{a'}(\vec{r}-\vec{r}') \tau_{\rm 
ind}(\vec{r}-\vec{r}') {\rm d}^2r'\label{eq:taufp}.
\end{eqnarray}

Since $d_{++}$ and $d_{--}$ are even functions in Eqs. (\ref{eq:tauf},\ref{eq:taufp}) for nearly 
homogeneous systems the contribution of the difference $d_{++}-d_{--}$ to $\tau_f(\vec{r})$ and $ 
\tau'_f(\vec{r})$ can be neglected resulting in
\begin{eqnarray}
 \tau'_f(\vec{r})=\frac{\kappa(\vec{r})}{\rho(\vec{r})}\tau_f(\vec{r}). \label{eq:tau_p}
\end{eqnarray}
From Eqs. (\ref{eq:taudv},\ref{eq:tauda},\ref{eq:tau_p}) one gets
\begin{eqnarray}
 \tau_v=-\tau_f+\tau_b \label{eq:tau_fb} 
\end{eqnarray}
and
\begin{eqnarray}
\tau_a= \frac{\kappa}{\rho}\tau_f+\tau_d \label{eq:tau_fd} 
\end{eqnarray}
After substituting Eqs. (\ref{eq:tau_fb},\ref{eq:tauda}) into Eqs. 
(\ref{eq:rhodd+},\ref{eq:rhodd-}) one concludes 
\begin{eqnarray}
 &&  \partial_t \rho_+(\vec{r},t)= \label{eq:rhod2+} \nonumber\\&&
-M_0b\partial_x 
\left\{\rho_+\left[ \tau_{\rm mf}+\tau_b-\left(1-\frac{\kappa}{\rho}\right)\tau_f+\tau_d 
\right]\right\} 
\end{eqnarray}
\begin{eqnarray}
 &&   \partial_t \rho_-(\vec{r},t)= \label{eq:rhod2-} \nonumber\\&&
+M_0b\partial_x 
\left\{\rho_-\left[ \tau_{\rm mf}+\tau_b-\left(1+\frac{\kappa}{\rho}\right)\tau_f-\tau_d 
\right]\right\} 
\end{eqnarray}
with $\tau_{\rm mf}=\tau_{\rm ext}+\tau_{\rm sc}$.

By adding and subtracting the above equations we obtain
\begin{eqnarray}
 &&  \partial_t \rho(\vec{r},t)= \label{eq:rhod2} \nonumber\\&&
-M_0b\partial_x 
\left[ \kappa\tau_{\rm mf}+\kappa\tau_b+\rho\tau_d 
\right] \\ 
 &&   \partial_t \kappa(\vec{r},t)= \label{eq:kappa2} \nonumber\\&&
-M_0b\partial_x 
\left[ \rho\tau_{\rm mf}+\rho\tau_b-\rho\tau_f +\frac{\kappa^2}{\rho}\tau_f+\kappa\tau_d 
\right]. 
\end{eqnarray}

Since according to DDD simulations \cite{groma1998,groma2003} and theoretical arguments 
\cite{groma2007,zaiser2015}, the correlation functions decay to zero faster than algebraically
on scales $|x - x'| \gg 1/\sqrt{\rho}$, in the above expressions for $\tau_v$ and $\tau_a$ the 
densities $\rho(\vec{r}')$ and $\kappa(\vec{r}')$ can be approximated by their Taylor expansion around the 
point $\vec{r}$. Assuming that the spatial derivatives of the densities are small on the scale 
of the mean dislocation spacing, $\partial_x \kappa/\kappa \ll \sqrt{\rho}, \partial_x \rho/\rho \ll 
\sqrt{\rho}$, we can retain only the lowest-order nonvanishing terms \cite{groma2003}. Since $\tau_{\rm 
ind}(x,y)=-\tau_{\rm ind}(-x,y)$ and $\tau_{\rm ind}(x,y)=\tau_{\rm ind}(x,-y)$, from the symmetry properties of the correlation 
functions mentioned above one concludes that up to second order
\begin{eqnarray}
\tau_f(\vec{r})&=&-\sqrt{\rho(\vec{r})} \int d_a(\vec{\tilde{r}})\tau_{\rm 
ind}(\vec{\tilde{r}}){\rm d}^2\tilde{r} \nonumber \\ 
&-& \frac{\partial_{xx}\rho(\vec{r})}{\rho^{3/2}} \int d_a(\vec{\tilde{r}}) \tilde{x}^2\tau_{\rm 
ind}(\vec{\tilde{r}}){\rm 
d}^2\tilde{r} \nonumber \\ 
&=&-\mu b C \sqrt{\rho(\vec{r})} \left(1 + \frac{\eta}{\rho^2} \partial_{xx} \rho(\vec{r})\right), 
\label{eq:flowstress}
\end{eqnarray}
\begin{eqnarray}
\tau_b(\vec{r})&=&-\partial_x\kappa(\vec{r})\int \tilde{x} d_s(\vec{\tilde{r}})\tau_{\rm 
ind}(\vec{\tilde{r}}){\rm 
d}^2\tilde{r} \nonumber \\
&=&-Gb \frac{D}{\rho}\partial_x\kappa(\vec{r}), \label{eq:backstress}
\end{eqnarray}
where $\vec{\tilde{r}}= \sqrt{\rho}\vec{r}$, $\tilde{x}= \sqrt{\rho}x$, and 
$G=\frac{\mu}{2\pi(1-\nu)}$.  
With the same notations we find
\begin{eqnarray}
\tau_d(\vec{r})&=&-\partial_x\rho(\vec{r})\int \tilde{x} d_p(\vec{\tilde{r}})\tau_{\rm 
ind}(\vec{\tilde{r}}){\rm 
d}^2\tilde{r} \nonumber \\
&=&-Gb \frac{A}{\rho}\partial_x\rho(\vec{r}).  \label{eq:dstress}
\end{eqnarray}
With Eqs.~(\ref{eq:tau_fb},\ref{eq:backstress},\ref{eq:tau_fd},\ref{eq:dstress}) the evolution 
equations 
(\ref{eq:rhod2},\ref{eq:rhod2}) read
\begin{eqnarray}
\label{eq:rhof}
& \partial_t \rho = - M_0 b\partial_x &
\left\{\kappa\tau_{\rm mf} -GbD\frac{\kappa}{\rho}\partial_x\kappa-Gb 
A\partial_x\rho\right\},\nonumber\\
\\
& \partial_t \kappa \label{eq:kappaf} = - M_0b\partial_x & \left\{\rho\left[ 
\tau_{\rm mf}-\left(1-\frac{\kappa^2}{\rho^2}\right)\tau_{\rm f}\right]\right.
\nonumber\\
&& \left. -GbD\partial_x\kappa -GbA\frac{\kappa}{\rho}
\partial_x\rho \right\}.
\end{eqnarray}

It is important to point out that in general the correlation functions are stress dependent. As a 
consequence, the parameters $C$, $D$, and $A$ introduced above can depend on the long-range 
stress $\tau_{\rm mf}$, which is in general a non-local functional of the 
excess dislocation density $\kappa$. More precisely, from dimensionality reasons it follows that 
parameters may depend on the dimensionless parameter $\tau_{\rm mf}/(\mu b \sqrt{\rho})$. From 
the symmetry properties of the correlation 
function one can easily see that  $C$ is an odd, $D$, and $A$ are even functions of $\tau_{\rm 
mf}$. As a consequence, at $\tau_{\rm mf}=0$, $C$ vanishes, while $D$, and $A$ have finite values 
and so they can be approximated up to second order in $\tau_{\rm mf}$ by constants. 

To establish the stress dependence of the parameter $C$ we note that
due to the relation (see Ref. \cite{groma2003})
\begin{eqnarray}
 \partial_t \kappa(\vec{r},t)=- \frac{1}{b} \partial_x \dot{\gamma}(\vec{r},t),
\end{eqnarray}
an explicit expression for the plastic shear rate in a homogeneous system is given by 
\begin{equation}
\dot{\gamma}= \rho b M_0 \left[\tau_{\rm mf} - \left(1 - \frac{\kappa^2}{\rho^2}\right) \tau_{\rm 
f}\right].
\end{equation}
where $\tau_{\rm f} = \mu b C \sqrt{\rho}$. 
If we consider a system without excess dislocations, such a system exhibits a finite flow stress due to
formation of dislocation dipoles or multipoles. For stresses below the flow stress, the strain rate is zero. 
It must therefore be
\begin{equation}
C = \left\{ \begin{array}{ll}
\alpha \frac{\tau_{\rm mf}}{\mu b \sqrt{\rho} }, \quad & |\tau_{\rm mf}| < \alpha \mu b \sqrt{\rho}\\
\alpha ,\quad&  |\tau_{\rm mf}| \ge \alpha \mu b \sqrt{\rho}
\end{array}
\right.
\label{eq:C}
\end{equation}
Here the former case corresponds to stresses below the flow stress, and the latter case to stresses above the flow stress. 
In a system where excess dislocations are present, the excess dislocations cannot be pinned by dipole/multipole formation 
but their effective mobility is strongly reduced. Only in the limit $\kappa = \rho$ the 
effective mobility of the dislocations reaches the value $M_0$ of the free dislocation. 
  
In conclusion we note that apart from the actual form of $\tau_d$ the 
evolution equations were derived earlier \cite{groma2003,groma2007}. The possible importance of 
$\tau_d$ has been recently raised by Finel and Valdenaire \cite{finel,valdenaire}. We also note 
that, for a rather special 
dislocation configuration where dislocations are artificially placed on periodically arranged
slip planes, a diffusion like term proportional to $\partial_x \rho$ has been derived by 
Dogge {\it et. al.} \cite{dogge2015}. However, in this type of analysis an artificial length scale 
parameter is introduced in terms of the slip plane spacing and, as demonstrated elsewhere 
\cite{zaiser2013}, the results depend crucially on the difficult-to-justify assumption of a 
strictly periodic arrangement of active slip planes. 

Since the parameters $D$ and $A$ are directly related to the correlation function, it should be 
possible to determine their actual values from correlation functions obtained by DDD simulations. 
For reasons that will be discussed elsewhere, this is however difficult unless analytical 
approximations for these functions are known, and indirect methods are more reliable. Thus, for 
nontrivial systems like a slip channel under load \cite{groma2003}, the dislocation configuration 
around hard inclusions, \cite{groma2004}, and the induced excess dislocations surrounding any 
given dislocation in a 
dislocation system \cite{groma2006,zaiser2015}, the parameter $D$ has been determined by direct 
comparison of DDD simulation results and solutions of the continuum equations. $D$ is found to 
be in the range of 0.25 to 0.8, while $A$ is found to be about 1.25 by Valdenaire 
\cite{valdenaire}.   

\section{Variational approach}

We have shown in earlier work that the evolution equations for the two densities of positive and 
negative dislocations
as studied above can be cast into the framework of phase field theories 
\cite{groma2006,groma2007,groma2010,groma2015}. 
The terms proportional to $\tau_a$, however, were not included into the earlier considerations. We 
now demonstrate that 
these terms can be equally obtained from an appropriate energy functional using the phase field 
formalism.    

For a system of straight parallel edge dislocations with Burgers vectors parallel to the $x$ axis
the evolution equations of dislocation densities $\rho_+$ and $\rho_-$ have the 
form \cite{groma2007,dogge2015}
\begin{eqnarray}
 \partial_t\rho_{\pm}+\partial_x[\rho_{\pm} v_{\pm}]=\pm f(\rho_+,\rho_-) \label{eq:cons}
\end{eqnarray}
in which we consider only dislocation glide, climb is neglected. Here $ v{\pm}$ is the glide 
velocity of 
positive or negative signed dislocations, and $f(\rho_+,\rho_-)$ is a term accounting for 
dislocation multiplication or annihilation. Since multiplication terms cannot be derived for 
2D systems (straight dislocations cannot multiply) but need to be introduced via ad-hoc 
assumptions, we assume that the number of dislocations is conserved, {\it i.e.}, we consider 
the limit $f(\rho_+,\rho_-)=0$ and focus on the $\rho_{\pm}$ dependence of the velocities  $ 
v_{\pm}$.

We adopt the standard formalism of phase field theories of conserved quantities. Assuming 
proportionality between fluxes and driving forces we have
\begin{eqnarray}
 v_+&=&-M_0\left\{\partial_x\left[\frac{1+\zeta}{2}\frac{\delta P}{\delta 
\rho_+}-\frac{1-\zeta}{2}\frac{\delta P}{\delta \rho_-}\right]\right\} \label{eq:v+} \\
v_-&=&-M_0\left\{\partial_x\left[\frac{1+\zeta}{2}\frac{\delta P}{\delta 
\rho_-}-\frac{1-\zeta}{2}\frac{\delta P}{\delta \rho_+}\right]\right\} \label{eq:v-}
\end{eqnarray}
where $P[\rho_+,\rho_-]$ is the phase field functional and $\zeta$ is 
a parameter. From Eqs. (\ref{eq:cons},\ref{eq:v+},\ref{eq:v-}) the evolution equations for the dislocation 
densities derive as
\begin{eqnarray}
 \partial_t \rho_+-\partial_x \left\{\rho_+ M_0\left[\partial_x \frac{\delta P}{\delta \kappa}+ 
\zeta\partial_x \frac{\delta P}{\delta \rho} \right]\right\}&=&0, \\
\partial_t \rho_-+\partial_x \left\{\rho_- M_0\left[\partial_x \frac{\delta P}{\delta \kappa}- 
\zeta\partial_x \frac{\delta P}{\delta \rho} \right]\right\}&=&0.
\end{eqnarray}
Accordingly we find
\begin{eqnarray}
\partial_t \rho &=& \partial_x \left\{\kappa M_0 \partial_x \frac{\delta P}{\delta \kappa}+ \zeta 
\rho 
 M_0 \partial_x \frac{\delta P}{\delta \rho}\right\}, \label{eq:PPrho}\\
\partial_t \kappa&=&\partial_x \left\{\rho M_0\partial_x \frac{\delta P}{\delta \kappa}+ 
\zeta \kappa M_0 \partial_x \frac{\delta P}{\delta \rho}\right\} \label{eq:PPkappe}.
\end{eqnarray}
In previous derivations \cite{groma2007} the terms proportional to $\delta P/\delta \rho$ were 
not considered ($\zeta = 0$). As discussed below, these terms are closely related to $\tau_d$ 
introduced above.   
Concerning the actual form of $P[\rho_+,\rho_-]$ it is useful to split it into two parts, the ``mean 
field'' or ``self consistent''  part $P_{\rm sc}$ and the ``correlation'' part  $P_{\rm c}$ 
which are defined below.  

In order to obtain the equation for the mean field stress $\tau_{\rm mf}$  from a variational 
principle for $P_{\rm sc}$ we represent the associated elastic energy using the Airy stress 
function formalism. By taking 
\begin{eqnarray}
 P_{\rm sc}[\chi,\rho_+,\rho_-]=\int \left[-\frac{1-\nu}{4\mu}(\bigtriangleup\chi)^2+b\chi 
\partial_y\kappa\right] {\rm d}^2r \label{eq:P_sc}
\end{eqnarray}
the minimum condition
\begin{eqnarray}
 \frac{\delta P_{\rm sc}}{\delta \chi}=0 \label{eq:chi_min}
\end{eqnarray}
leads to the equation
\begin{eqnarray}
 \frac{1-\nu}{2\mu}\bigtriangleup^2\chi=b\partial_y\kappa \label{eq:chi}
\end{eqnarray}
where $\chi$ is the Airy stress function from which the shear stress derives via $\tau_{\rm 
mf}=\partial_x\partial_y \chi$. The general solution of Eq.~(\ref{eq:chi}) is $\tau_{\rm mf}$ given 
by Eq.~(\ref{eq:tausc}) plus the external stress. Substituting Eq.~(\ref{eq:P_sc}) into 
(\ref{eq:v-}) and 
(\ref{eq:v+}) one gets 
\begin{eqnarray}
   \partial_t \rho_+(\vec{r},t) + M_0b\partial_x 
\left(\rho_+ \tau_{\rm mf} \right)=0,\\
   \partial_t \rho_-(\vec{r},t) - M_0b\partial_x 
\left(\rho_- \tau_{\rm mf}\right)=0.
\end{eqnarray}
$P_{\rm sc}$ merely recovers the mean field part of the dislocation velocities $v_+$ and 
$v_-$ but not the terms which are related to dislocation-dislocation 
correlations. It thus needs to be complemented by a 'correlation' part 
of the phase field functional. We use a form which can be derived by means of a similar 
averaging strategy as used above for the driving forces, but applied to the elastic
energy functional of the discrete dislocation system \cite{zaiser2015}. For the present dislocation
system the resulting `correlation' part of the phase field functional is given by 
\begin{equation}
   P_{\rm corr}=\int\left[ Gb^2A \rho\ln\left(\frac{\rho}{\rho_0}\right)
  +\frac{Gb^2 D}{2}\frac{\kappa^2}{\rho}\right]{\rm d }^2 r \label{eq:p_corr}.
\end{equation}
This expression is tantamount to using a local density approximation for the correlation 
part (the functional contains only on the local values of the dislocation densities, 
not any gradients or non-local expressions). 

We consider weakly polarized dislocation arrangements where
$\kappa/\rho \ll 1$ and $\partial_x \rho/\rho \ll \rho^{1/2}$. By neglecting terms of higher than 
first order in $\kappa/\rho$ and $\partial_x \rho/\rho^{3/2}$, 
we find that
\begin{eqnarray}
&& \partial_t \rho_+ = \label{eq:Prhof+}
\nonumber\\&&-\partial_x 
\left[\rho_+M_0 b\left( \tau_{\rm mf}-Gb\frac{D}{\rho}\partial_x\kappa-Gb 
\zeta\frac{A}{\rho}\partial_x\rho \right)\right],\\
&& \partial_t \rho_- = \label{eq:Prhof-}
\nonumber\\&&+\partial_x 
\left[\rho_-M_0b\left( \tau_{\rm mf}-Gb\frac{D}{\rho}\partial_x\kappa+Gb 
\zeta\frac{A}{\rho}\partial_x\rho \right)\right]. 
\end{eqnarray}
From the above equations the evolution equations for $\kappa$ and $\rho$ read 
\begin{eqnarray}
\partial_t \rho= \label{eq:Prhof}
-M_0b\partial_x 
\left\{\kappa\tau_{\rm mf} -GbD\frac{\kappa}{\rho}\partial_x\kappa-Gb 
A\partial_x\rho\right\},
\end{eqnarray}
\begin{eqnarray}
\partial_t \kappa= \label{eq:Pkappaf}  
-M_0b\partial_x 
\left\{\rho
\tau_{\rm mf}-GbD\partial_x\kappa-GbA\frac{\kappa}{\rho}
\partial_x\rho \right\}.
\end{eqnarray}
With $\zeta=1$, apart from the term containing the ``friction'' stress $\tau_{\rm f}$,  Eqs. 
(\ref{eq:Prhof}, \ref{eq:Pkappaf}) are equivalent to Eqs. (\ref{eq:rhof}, 
\ref{eq:kappaf}). So, with the appropriate form of the correlation term in the phase field 
functional (a form which derives from ensemble averaging the energy functional of the discrete dislocation 
system), by applying the standard formalism of phase field theories we recover 
the evolution equations of the dislocation densities derived by ensemble averaging 
the equations of motion of individual dislocations. However, the flow stress which plays 
a crucial role in plastic deformation of any material can {\em not} be directly derived 
within the traditional framework of phase field 
theories.  

To resolve this issue we modify Eq. (\ref{eq:PPkappe}) to allow for a non-linear dependency on 
the driving force $\delta P/\delta \kappa$. The modified equation is given by
\begin{eqnarray}
 \partial_t \kappa&=&\partial_x \left\{\rho M\left(\partial_x \frac{\delta P}{\delta 
\kappa}\right)+\kappa M_0 \partial_x \frac{\delta P}{\delta \rho}\right\}
\label{eq:PPMkappa}
\end{eqnarray}
were $M(x)$ is a nontrivial mobility function defined as
\begin{equation}
 M(x)=M_0\left\{ 
\begin{array}{ll}
      \frac{\kappa^2}{\rho^2}x \ \ & {\rm if} \ \ |x|<x_0 \\
      {\rm sgn}(x) \left[ |x| - x_0 \left(1-\frac{\kappa^2}{\rho^2} \right)\right] \\ & {\rm if} \ \ 
|x|\geqq x_0
    \end{array}
 \right.
\label{eq:Mpm}
\end{equation}
with $x_0=\alpha\mu b^2\sqrt{\rho}$ (see fig.1). 
\begin{figure}[!ht]
\includegraphics[angle=0,width=6cm]{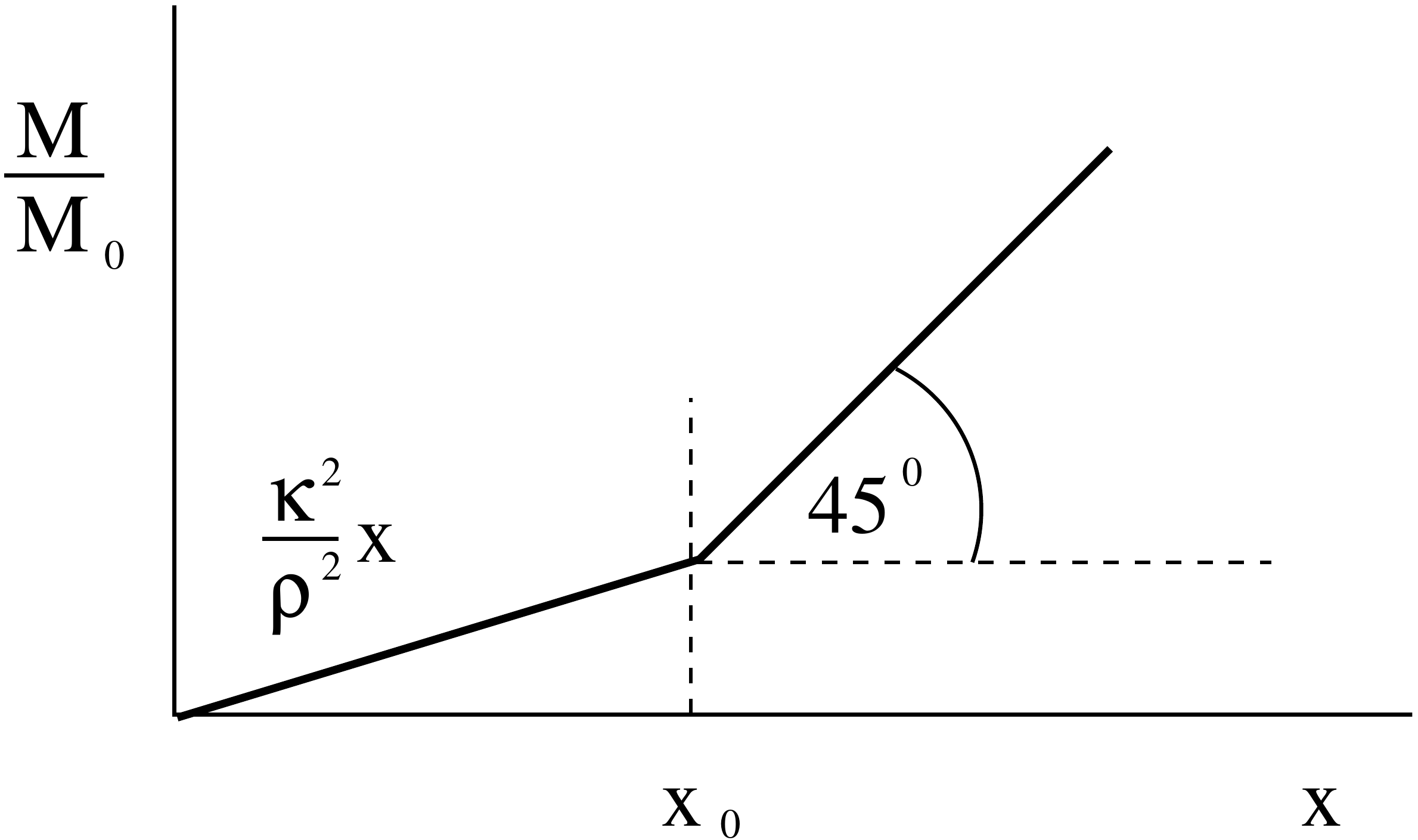}
\caption{The $M(x)$ mobility function}
\end{figure}
It is easy to see that this mobility function recovers Eq.~(\ref{eq:kappaf}) within the 
general framework of a phase field theory. 

As concluding remarks of this discussion it should be noted that: 
\begin{itemize}
 \item Since thermal energies are typically 4-5 orders of 
magnitude lower than the elastic energies associated with the presence of dislocations 
\cite{zaiser2015}, entropic contributions to the dislocation free energy are negligible 
up to the melting point. The requirement of thermodynamic consistency of any theory 
in this case reduces to the trivial requirement that the elastic energy must decrease 
and can never increase during system evolution (the latter would imply a transfer of 
energy from the heat bath to the elastic energy of the crystal). The comparison of 
the evolution equations obtained by direct averaging and the phase field formalism 
indicates that the evolution equations of the dislocation densities can nevertheless be 
cast into the phase field framework. 

\item The irrelevance of thermal fluctuations 
makes it mandatory to introduce a nontrivial on/off type mobility function. The reason is
that dislocations, as they move through the crystal, not only experience the {\em average
energy} expressed by the functional $P$, but also {\em energy fluctuations} on scales comparable
to the dislocation spacing. The magnitude of these fluctuations scales like 
$\alpha G b \sqrt{\rho}$, as discussed e.g. by Zaiser and Moretti \cite{zaiser2005}. 
Since thermal fluctuations of sufficient magnitude
are not available, the work required to overcome these fluctuations and to enable sustained
dislocation motion must be provided by the local stress. This is reflected by the mobility functions 
$M(x)$ which introduce a contribution akin to dry friction into the dynamics - for the derivation 
of a similar friction-like stress contribution in 3D  (see \cite{zaiser2015}).  

\item The forms of the phase field functional and the mobility functions suggested here represent 
only the simplest possible approximation, which is correct for weakly polarized and weakly inhomogeneous
dislocation arrangements only. For some specific problems like dislocation distribution next to boundaries, 
or strongly inhomogeneous systems, one may have to consider additional terms (see e.g. Ref. 
\cite{groma2015}).
\end{itemize}

\section{Time variation of the phase field functional}

The time derivative of the phase field functional is
\begin{eqnarray}
 \frac{dP}{dt}=\frac{\delta P}{\delta \rho}\partial_t \rho + \frac{\delta P}{\delta 
\kappa}\partial_t \kappa +\frac{\delta P}{\delta \chi}\partial_t \chi
\end{eqnarray}
but due to the condition (\ref{eq:chi_min}) the third term vanishes. For simplicity in the following 
the phase field functional is always evaluated at $\chi=\chi_{min}$ 
defined by $\delta P/\delta \chi |_{\chi_{min}}=0$.
Hence 
\begin{eqnarray}
 \frac{dP}{dt}=\left. \left( \frac{\delta P}{\delta \rho}\partial_t \rho + \frac{\delta P}{\delta 
\kappa}\partial_t \kappa \right) \right |_{\chi=\chi_{min}} \label{eq:dpdt}.
\end{eqnarray}
By substituting Eqs. (\ref{eq:PPrho},\ref{eq:PPMkappa}) into Eq. (\ref{eq:dpdt}) we obtain after 
partial integration
\begin{eqnarray}
 \frac{dP}{dt}&=&-\left\{\partial_x \frac{\delta P}{\delta \rho}\right\}
 \left\{\kappa M_0 \partial_x \frac{\delta P}{\delta \kappa}+\rho 
 M_0 \partial_x \frac{\delta P}{\delta \rho}\right\}
\nonumber \\ &&-
\left\{\partial_x \frac{\delta P}{\delta\kappa}\right\}
\left\{\rho M\left[\partial_x \frac{\delta P}{\delta 
\kappa}\right]+\kappa M_0 \partial_x \frac{\delta P}{\delta \rho}\right\}.
\label{eq:dpv}
\end{eqnarray}
If $|\partial_x \delta P/\delta \kappa|<\alpha\mu b^2\sqrt{\rho}$
\begin{eqnarray}
  \frac{dP}{dt}=-M_0\left(
\begin{array}{l}
       \partial_x \frac{\delta P}{\delta \rho} \\
       \partial_x \frac{\delta P}{\delta \kappa}
    \end{array}
 \right)
 \left(\begin{array}{ll}
        \rho , & \kappa \\
        \kappa, & \kappa^2/\rho
       \end{array}
\right)
\left(
\begin{array}{l}
       \partial_x \frac{\delta P}{\delta \rho} \\
       \partial_x \frac{\delta P}{\delta \kappa}
    \end{array}
 \right).
\end{eqnarray}
Since $M_0$ is positive and the matrix
\begin{eqnarray}
  \left(\begin{array}{ll}
        \rho, & \kappa \\
         \kappa, & \kappa^2/\rho
       \end{array}
\right)
\end{eqnarray}
is positive definite it follows that ${dP}/{dt}\leq 0$. 
In the flowing regime ($|\partial_x \delta P/\delta \kappa|>b \alpha\mu b^2\sqrt{\rho}$)
we find that
\begin{eqnarray}
&&  \frac{dP}{dt}=-M_0\left(
\begin{array}{l}
       \partial_x \frac{\delta P}{\delta \rho} \\
       \partial_x \frac{\delta P}{\delta \kappa}
    \end{array}
 \right)
 \left(\begin{array}{ll}
         \rho , & \kappa \\
         \kappa, & 0
       \end{array}
\right)
\left(
\begin{array}{l}
       \partial_x \frac{\delta P}{\delta \rho} \\
       \partial_x \frac{\delta P}{\delta \kappa}
    \end{array}
 \right) \\ \nonumber
 &-& M_0  \rho \left(\partial_x \frac{\delta 
P}{\delta \kappa}\right)\;{\rm sgn}\left(\partial_x \frac{\delta 
P}{\delta \kappa}\right) \times\nonumber\\
&& 
\left[\left|\partial_x \frac{\delta 
P}{\delta \kappa}\right| - \alpha\mu b^2\sqrt{\rho} \left(1-\frac{\kappa^2}{\rho^2}\right)\right] 
\end{eqnarray}
This again ensures that ${dP}/{dt}\leq 0$. So, we found in both cases that the phase field 
functional cannot increase during the evolution of the system. Since our phase field functional is 
tantamount to the averaged elastic energy functional, this ensures thermodynamic consistency of our 
theory. 

\section{Pattern formation} 

In the following we discuss under what conditions the evolution equations derived above can lead to 
instability resulting in dislocation pattern formation. One can easily see that the trivial 
homogeneous solution $\rho=\rho_0$, $\kappa=0$ and $\tau_{\rm mf}=\tau_0$ satisfies Eqs. 
(\ref{eq:rhof},\ref{eq:kappaf},\ref{eq:chi}), where $\rho_0$ 
and $\tau_0$ are constants representing the initial dislocation density and the external shear 
stress, respectively. The stability of the trivial solution can be analyzed by applying the 
standard method of linear stability analysis. One can easily see that nontrivial behavior can happen only 
in the flowing regime {\it i.e.} if $|\tau_0| > \alpha \mu b \sqrt{\rho_0}$, so we consider only 
this case. 

By adding small perturbations to the dislocation densities and the Airy stress function in the form 
\begin{eqnarray}
 \rho(\vec{r},t)&=&\rho+\delta\rho(\vec{r},t) \nonumber \\
 \kappa(\vec{r},t)&=&\delta\kappa(\vec{r},t) \label{eq:perturbation} \\
 \chi(\vec{r},t)&=&\tau_0 xy+\delta \chi(\vec{r},t) \nonumber
\end{eqnarray}
and keeping only the leading terms in the perturbations, equations 
(\ref{eq:rhof},\ref{eq:kappaf},\ref{eq:chi}) become
\begin{eqnarray}
\partial_t \delta\rho &=& \label{eq:lrho}
M_0\partial_x\left[GbA\partial_x\delta\rho - \tau_0 \delta \kappa \right]
\end{eqnarray}
\begin{eqnarray}
\partial_t \delta\kappa &=& \label{eq:lkappa}
-M_0\Theta_f\partial_x 
\left[\rho_0\partial_x\partial_y \delta\chi-GbD\partial_x\delta 
\kappa\right] \\ 
&&-  M_0\Theta_f\left[\tau^* -  \alpha \mu b \frac{\sqrt{\rho_0}}{2} \right] 
\partial_x\delta\rho.
\nonumber
\end{eqnarray}
\begin{eqnarray}
\bigtriangleup^2\delta\chi=4\pi Gb\partial_y\delta\kappa \label{eq:lchi}
\end{eqnarray}
In these expressions, $\tau^* = \tau_0 - \alpha \mu b \sqrt{\rho_0}$, and the step function 
$\Theta_f = \Theta(\tau^*)$ is zero if the applied stress is below the flow stress in the 
homogeneous reference state, and 1 otherwise. To obtain the above equations it was taken into 
account that the first-order variation of the flow stress is 
given by
\begin{eqnarray}
\delta\tau_{\rm f} = \frac{\alpha \mu b \sqrt{\rho_0}}{2}
\frac{\delta \rho}{\rho_0}.
\label{eq:deltatauf}
\end{eqnarray}

The solution of Eqs. (\ref{eq:lchi},\ref{eq:lrho},\ref{eq:lkappa}) can be found in the form
\begin{eqnarray}
 \left( \begin{array}{ccc}
          \delta\rho \\
          \delta\kappa \\
          \delta\chi
        \end{array} \right)=
 \left( \begin{array}{ccc}
          \delta\rho_0 \\
          \delta\kappa_0 \\
          \delta\chi_0
        \end{array} \right) \exp\left(\frac{\lambda}{t_0}t+i\sqrt{\rho_0}\vec{k}\vec{r} \right)
        \label{eq:solution}
\end{eqnarray}
were $\vec{k}$ is a dimensionless quantity.
After substituting the above form into Eqs. (\ref{eq:lchi},\ref{eq:lrho},\ref{eq:lkappa}) in the
flowing regime  ($\Theta_f=1$)  we get
\begin{eqnarray} 
  \left(\begin{array}{cc}
         \lambda+Ak_x^2,  & i(\dot{\gamma}'+ 2 \alpha')k_x \\
         i (\dot{\gamma}'-\alpha')k_x, & \lambda+Dk_x^2+T(\vec{k})
     \end{array}\right)
     \left( \begin{array}{cc}
          \delta\rho \\
          \delta\kappa 
        \end{array} \right)=0 \label{eq:det}
\end{eqnarray}
where the notations $t_0=b^2G\rho_0/B$, $T(\vec{k})=
4\pi k_x^2k_y^2/|\vec{k}|^4$, $\dot{\gamma}'=\tau_*/(Gb\sqrt{\rho_0})$, and 
$\alpha'= \pi(1-\nu)\alpha$ were introduced. Note that in the above equations 
each of the parameters are dimensionless and $\dot{\gamma}'$ is proportional 
to the average shear rate $\dot{\gamma}=M_0b^2 \rho_0 \tau^*$.

Eq. (\ref{eq:det}) has nontrivial solutions if 
\begin{eqnarray} 
 (\lambda+Ak_x^2)(\lambda+Dk_x^2+T(\vec{k}))+k_x^2 \beta =0
\end{eqnarray}
with $\beta=(\dot{\gamma}'+2\alpha')(\dot{\gamma}'-\alpha')$. 
This leads to 
\begin{eqnarray}
 &&\lambda_{\pm}=-\frac{(A+D)k_x^2+T(\vec{k})}{2} \nonumber\\
&&\pm\frac{\sqrt{[(D+A)k_x^2+T(\vec{k})]^2-4k_x^2[\beta+A(Dk_x^2+T(
\vec{k}))]} }{ 2 }.\nonumber\\
\end{eqnarray}
It follows that the condition for the existence of growing perturbations ($\lambda>0$) is
\begin{eqnarray}
[\beta + AT(\vec{k})+ADk_x^2]<0.
\end{eqnarray}
$T(\vec{k})$ cannot be negative and it vanishes if $\vec{k}$ is parallel to either the $x$ or 
to the $y$ axis. Thus, $\beta < 0$ is a necessary and sufficient condition for instability. 
This condition requires that (i) the system is in the flowing phase and (ii) 
$\dot{\gamma}'$ must be smaller than $\alpha'$. In this case there exists 
a region in the $\vec{k}$ space in which  perturbations grow. 
Perturbations with wave vectors outside this region decay in time (see figs. 2,3). 
\begin{figure}[!ht]
\includegraphics[angle=0,width=7cm]{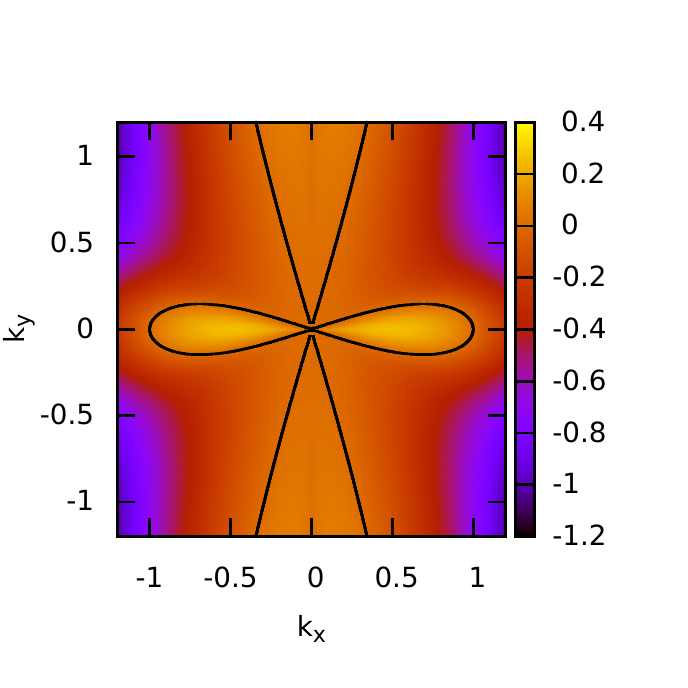}
\caption{The $\lambda_+(k_x,k_y)$ function at $A=1$, $D=1$ and $\beta =-1$. 
The function is positive within the region marked by 
the contour line $\lambda_+(k_x,k_y)=0$.}
\end{figure}
This results in a length scale selection corresponding to the fastest growing periodic perturbation 
$\vec{k}_{\rm max}$ defined by the condition 
\begin{eqnarray}
 \left. \frac{d \lambda_+(\vec{k})}{d\vec{k}}\right |_{\vec{k}_{\rm max}}=0
\end{eqnarray}

\begin{figure}[!ht]
\includegraphics[angle=0,width=6cm]{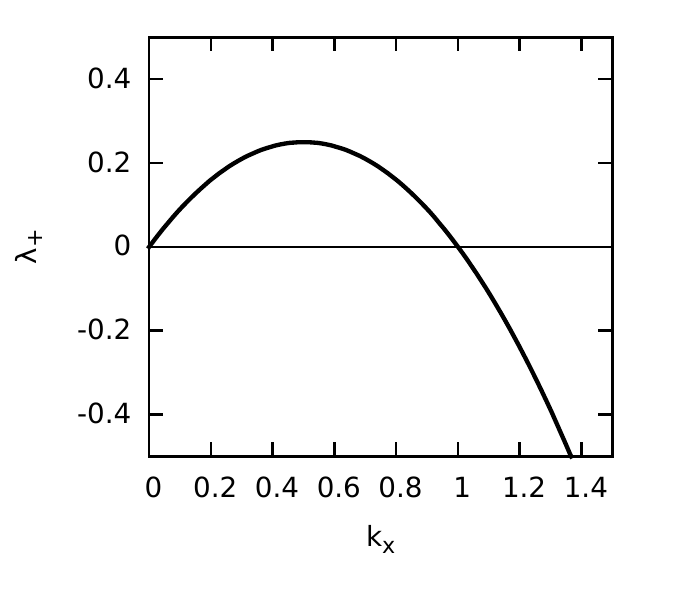}
\caption{The $\lambda_+(k_x,0)$ function at $A=1$, $D=1$ and $\beta =-1$.}
\end{figure}

For negative $\beta$, the $\lambda_+(k_x,k_y)$ function has two equal maxima along the $x$ axis 
located at 
\begin{eqnarray*}
 k_x^2= - 2 \beta \frac{-1+\sqrt{1+\frac{(A-D)^2}{4AD}}}{(A-D)^2} 
\end{eqnarray*}
It should be stressed that according to Eq. 
(\ref{eq:solution}) the actual wave vector of the fastest growing perturbation is $\sqrt{\rho_0} 
\vec{k}_{\rm max}$. So, in agreement with the principle of similitude observed experimentally the 
characteristic pattern  wavelength scales with the dislocation spacing $1/\sqrt{\rho_0}$. 
It is important to note at this point that the diffusion-like $\tau_d$ term introduced here plays a crucial role 
in characteristic wavelength selection. At $A=0$ (corresponding to $\tau_d=0$, see Eq. 
(\ref{eq:dstress})) perturbations of all wave vectors would grow and there would be no mode of 
maximum growth rate. In accordance with this, by analyzing the 
stability of the homogeneous solution of the 3D continuum theory of dislocations proposed by  
Hochrainer {\it et. al} \cite{hochrainer2007,hochrainer2014,hochrainer2015}, Sandfeld and Zaiser 
concluded \cite{sanfeld2015} that the mean field and the flow stresses generate instability but they 
do not result in length scale selection.       

Within the general framework introduced in the second section there is another way leading to 
dislocation pattern formation which can operate even if $A = 0$, but which requires the consideration 
of higher-order gradients in the dislocation densities. Until now we have neglected the term 
proportional to $\partial_{xx}\rho/\rho^{3/2}$ in the expression for the flow stress in Eq. 
(\ref{eq:flowstress}). Without going into the details of the derivation one can find that with 
this term, but with $A=0$, the evolution equations (\ref{eq:lrho}{,\ref{eq:lkappa}) 
get the form

 \begin{eqnarray}
\partial_t \delta\rho = \label{eq:lrho2}
-M_0 b\partial_x\left[\tau_0 \delta \kappa \right]
\end{eqnarray}  

\begin{eqnarray}
&&  \partial_t \delta\kappa = \label{eq:lkappa2}
- M_0 b\Theta_f \partial_x 
\left[\rho_0\partial_x\partial_y \delta\chi-GbD\partial_x\delta 
\kappa\right] \\ 
&&- M_0 b\Theta_f \left[\tau^* -\alpha \mu b \frac{\sqrt{\rho_0}}{2}  
\partial_x\delta\rho-\frac{Gb\eta}{\sqrt{\rho_0}} \partial_{xxx}\delta\rho\right], 
\nonumber
\end{eqnarray}  
where $\eta$ is a constant and for simplicity the terms related to $\tau_a$ are neglected. After 
substituting the solution given by Eq. (\ref{eq:solution}) into Eqs. 
(\ref{eq:lrho2}{,\ref{eq:lkappa2}) in the flowing regime ($\Theta_f=1$) we get 
\begin{eqnarray} 
  \left(\begin{array}{cc}
         \lambda,  & i (\dot{\gamma}'+ 2 \alpha')k_x \\
         ik_x[(\dot{\gamma}'-\alpha')+\eta k_x^2], & \lambda+Dk_x^2+T(\vec{k})
     \end{array}\right)
     \left( \begin{array}{cc}
          \delta\rho \\
          \delta\kappa 
        \end{array} \right)=0 \label{eq:det2}\nonumber\\
\end{eqnarray}
Eq. (\ref{eq:det2}) has nontrivial solutions if 
\begin{eqnarray} 
 \lambda(\lambda+Dk_x^2+T(\vec{k}))+k_x^2[(\dot{\gamma}'+ 2 
\alpha')(\dot{\gamma}'-\alpha'+\eta k_x^2)]=0 \nonumber\\
\end{eqnarray}
leading to
\begin{eqnarray}
 &&\lambda_{1,2}=-\frac{Dk_x^2+T(\vec{k})}{2} \\
&&\pm\frac{\sqrt{[Dk_x^2+T(\vec{k})]^2-4k_x^2[(\dot{\gamma}'+2 \alpha')(\dot{\gamma}'-\alpha'+\eta k_x^2)]} }{ 
2 }
\nonumber\\
\end{eqnarray}
The condition to growing perturbation is 
\begin{eqnarray}
(\dot{\gamma}'-\alpha')+\eta k_x^2<0
\end{eqnarray}
Provided that $\eta>0$ and $\dot{\gamma}'<\alpha'$ there is again a region in the $\vec{k}$ space 
in which  perturbations grow, and one can again find the wavelength corresponding to the fastest 
growing perturbation. It should be noted that in this case the length scale selection is caused by 
a second order effect in the sense that the term  $\partial_{xx}\rho/\rho^{3/2}$ is obtained by 
the second order Taylor expansion in Eq. (\ref{eq:taudv}) while $\tau_d$ given by Eq. 
(\ref{eq:dstress}) corresponds to a first order one.

\section{Conclusions}
In summary the general framework explained in detail in section II and III is able to account 
for the emergence of growing fluctuations in dislocation density leading to pattern 
formation in single slip. The primary source of instability is the $\sqrt{\rho}$ type of dependence 
of the flow stress, but alone it cannot lead to length scale selection. As it is shown above there 
are two alternative ways (the diffusion-like term associated with the stress $\tau_d$, or the 
$\partial_{xx}$ type term in the flow stress) leading to characteristic length scale of the 
dislocation patten. 

Irrespective of the pattern selection mechanism and in line with previous work \cite{sanfeld2015}, 
we find that there are two requirements for patterning: First, the system must be in 
the plastically deforming phase, second, the rate of shear must not be too high ($\dot{\gamma}' 
\le \alpha'$). This condition indicates that patterning as studied here can {\em not} be understood as a energy minimization
process, despite the fact that the dynamics which we investigate minimizes an energy functional. This seemingly 
paradox statement becomes clearer if we consider the limit $\alpha \to 0$ where the mobility functions become 
trivial. In this limit, the critical strain rate where patterning vanishes goes to zero. Thus, the patterning
is an effect of the non-trivial mobility function which introduces a strongly non-linear, dry-friction like 
behavior into the system. This aspect of the problem, which contradicts the low energy paradigm and emphasizes 
the dynamic nature of the patterning process, clearly should be further studied by extending the analysis into
the non-linear regime. 

In the limit of low strain rates, $\dot{\gamma} \ll \rho b M_0 \tau_{\rm ext}$, the selected 
pattern wavelength becomes independent on strain rate. The predictions in this regime agree well
with experimental observations: With $A = 1.25$ \cite{valdenaire}, $D =0.25$ \cite{groma2006}, 
$\alpha = 0.2$, and 
$\nu = 0.3$, we find a preferred wave-vector $|k_x| = \approx 0.42 \sqrt{\rho} $ corresponding to a wavelength of
about 15 dislocation spacings, in good agreement with typical observations. The preferred patterns corresponds to
dislocation walls perpendicular to the active slip plane, again in agreement with observations and discrete 
simulations. This  agreement does not mean that the present, very simple considerations alone provide a 
complete theory of dislocation patterning -- in particular the essential aspect of dislocation multiplication, 
and hence work hardening, is missing. However, it indicates that we may capture some of the essential features
of the real process. 

\begin{acknowledgments}
The authors are grateful to Prof. Alponse Finel for drawing their attention to the fact that the 
``asymmetric'' stress component neglected in earlier considerations may play a significant role. 
 Financial supports of the Hungarian Scientific Research Fund (OTKA) under
contract numbers K-105335  and PD-105256 and of the European Commission under grant agreement No. 
CIG-321842 are also acknowledged. PDI is supported by the J\'anos Bolyai Scholarship of the 
Hungarian Academy of Sciences.
\end{acknowledgments}


\end{document}